\documentclass[conference]{IEEEtran}
\usepackage{cite}
\usepackage{amsmath,amssymb,amsfonts}
\usepackage{algorithmic}
\usepackage{graphicx}
\usepackage{textcomp}

\def\BibTeX{{\rm B\kern-.05em{\sc i\kern-.025em b}\kern-.08em
    T\kern-.1667em\lower.7ex\hbox{E}\kern-.125emX}}
\begin{document}

\title{Design and Implementation of an Automatic Synchronizing and Protection Relay through Power-Hardware-in-the-Loop (PHIL) Simulation}

\author{\IEEEauthorblockN{Mishal Mahmood, Mariam Azam, Khair-un-Nisa Fatima, Muhammad Sarwar \\ Muhammad Abubakar, Babar Hussain}
\IEEEauthorblockA{\textit{Department of Electrical Engineering} \\
\textit{Pakistan Institute of Engineering and Applied Sciences}, Islamabad, Pakistan\\
\noindent \{mashal.mahmood.5,mariamazam997\}@gmail.com, khairunisa.fatima@yahoo.com, \{msarwar,abubakar\}@pieas.edu.pk}

%
%
%

}

\maketitle

\begin{abstract}
This paper focuses on the design and implementation of an automatic synchronizing and protection relay to automate the synchronization process of a Distributed Energy Resource (DER) to the Main Grid. The proposed design utilizes a cost-effective data acquisition using Arduino in combination with LabVIEW software to implement the multi-purpose synchronizing relay. The proposed synchronizing relay is capable of synchronizing a Distributed Generator (DG) to the power grid from black-start and fulfils the requirements imposed by the utility. The synchronizing relay is implemented through voltage and frequency control of an actual lab-scale synchronous generator. In the synchronization process, frequency synchronization is done using speed control of the stepper motor as the prime mover and voltage synchronization is accomplished using Excitation Control module through Power-Hardware-in-the-Loop (PHIL) simulation. In grid-connected mode, active and reactive power controls and protection schemes for the synchronous generator have also been implemented. The proposed multi-function relay has been deployed and tested on a lab-scale test bed to validate the proposed design and functionality.
\end{abstract}


\section{Introduction}

Historically, centralized power generating stations are being used as the major source of electric power. However, recently, environmental concerns and depleting fossil fuels have accelerated the role of renewable energy resources to fulfill energy demand. With the increasing energy demand, it has become necessary to deploy small scale distributed generators using different power generation sources near the consumer end. The electric grid expansion and need of installation of new power plants has become one of the most important topics of today. Distributed Generators (DG) can efficiently cater to the problem of ever increasing energy requirements by engaging small scale renewable energy sources near to the consumer end \cite{b1}.

Currently, there is a gradual paradigm shift to utilize different small scale power generation technologies. So, in modern power grids, DGs are widely deployed to decrease power losses and increase dependence on sustainable energy. As most of the small-scale power plants use synchronous generators, they need to be synchronized with power grid to transfer power to the loads \cite{b2}.

For the synchronization process, the connecting generator and the grid must have same:
\begin{enumerate}
\item Phase Sequence.
\item Voltage Magnitude.
\item Frequency.
\item Phase Angle.
\end{enumerate}

For the synchronization of a generator with main power stream, different countries have different grid specifications. An automatic synchronization should be able to adjust the parameters accordingly. In Pakistan, power grid has the standard
frequency of 50 Hz and voltage peak to peak of 400 V.

Poor synchronization causes disturbances in the power system as well as severe damages to the generator and transients in power system. It damages the prime mover and generator because of mechanical stresses caused by rapid acceleration or deceleration. It causes high currents that can cause damage to transformers, power lines and the generator. In the absence of protection schemes, these faults can spread in whole power system thus leading to a blackout.

IEEE Standards C50.12 and C50.13 provide specifications for generator synchronization \cite{b15}. 
If the breaker is closed without satisfying the conditions of synchronization, it can produce a short- circuit and can cause high vibrations from the torsional swaying of the pole. Protection of the DG against any failure in synchronization operation is neccessary \cite{b3}.

Auto-synchronization is finding expanded application in distribution frameworks. Experimentation is being done with different synchronization advances entirely automatic. Power organizations have begun putting resources into this innovation as they see potential in these methods.

A huge amount of work has been done earlier by engineers and researchers on the topic regarding the automatic synchronization of the synchronous generators. Iskandar Hack used LabVIEW, PCI-6014 Data Acquisition card, and the NI ELVIS devices to design a system that allows effortless and economical synchronization of small synchronous systems \cite{b18}. 

Erdal Bekiroglu proposed to use a microcontroller to control and automate the synchronization of two generators \cite{b19}. In his proposed technique, the data is evaluated by the algorithm coded into the microcontroller. The program needs to be updated for a new generator to get connected. Synchronization of DG with grid has not been discussed.

Nutthaka Chinomi used a data acquisition card and LabVIEW for a renewable supervisory and evaluating system. He states in his research paper that LabVIEW is a cheap alternative to measure the system parameters to typical analogue and digital measuring instruments. The data collected is associated with the data collected from the referenced device \cite{b20}. 

C. Navitha proposed a technique to use a PIC microcontroller, Reduced Instruction Set Computer (RISC) architecture and an internal Analog-to-Digital Converter (ADC) to get a system with a main focus to manage a solar-wind power system using LabVIEW. The simulation circuits are developed using elements of LabVIEW software \cite{b21}.

For the proposed automatic synchronization relay, LabVIEW is used to auto synchronize the generator to the grid system and to measure the synchronous generator’s steady state characteristics. This software is used to create a scalable control application which interacts with the real-world processes. It is a cost effective and efficient way for the implementation of auto-synchronization process.

The prototype is designed and testing using Power Hardware-In-the-Loop (PHIL). Voltage and current signals are processed by the software to generate actuator signals to control and synchronize the DG with grid. Frequency of the grid and the connecting source is also measured and speed of the prime mover of connecting source is adjusted according to the frequency of the grid.

The research work in this document covers the detail of automatic synchroniation method of any type or size of DG with grid. The system continuously monitors, protects and adjusts the control parameters of DG to synchronize with the grid in a reliable manner. The proposed system will potentially influence and enhance the Electrical Power Systems and Generation, by encouraging the use of DG. Thus, it will promote the use of renewables. Deploying DGs at the distribution side also cut the power losses in the transmission lines and thus improves the efficiency of system.

The second section of this document includes modelling of the power generation system.The third section includes the proposed design requirement, the details of design and its working. The fourth section describes the proposed protection scheme for the generator system and its details of working. The last section discusses the conclusions. At the end, a flowchart shows the summary of the proposed auto-synchronization scheme.

\section{Modelling of Electrical Power Generator System}
\begin{figure}[]
\centerline{\includegraphics[width=0.5\textwidth]{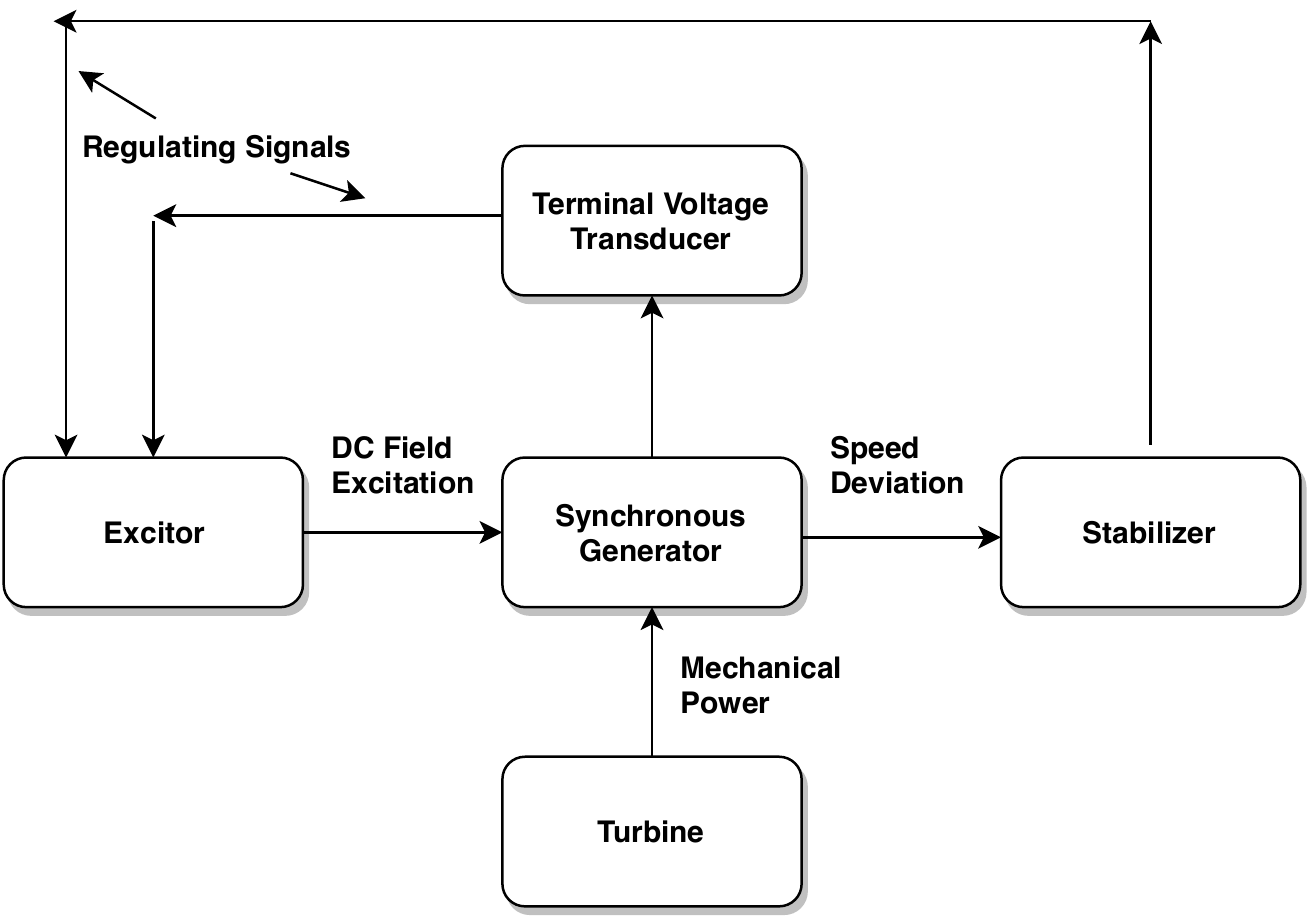}}
\caption{Modelling of power generator system}
\label{fig}
\end{figure}

Fig.~\ref{fig} shows a block diagram in which a synchronous generator is used to model the connecting source with the grid \cite{b4, b5}. It is modelled with its mechanical power provided by a prime mover and the field voltage provided by an exciter \cite{b6}. Generator rotor oscillations are damped using the Power System Stabilizer (PSS) \cite{b7, b8}. The system is provided Automatic Voltage Regulation (AVR) by controlling its field excitation system \cite{b9}. The frequency synchronization is done by controlling the speed of prime mover. Exciter receives the signals from generator terminal voltage transducer and the stabilizer\cite{b10, b11}. 

The generator used for the modelling of DG and for the experimentation purpose is a three phase 1 KW synchronous generator. Its specifications are shown in Table~\ref{tab1}.
\begin{table}[]
\caption{Specifications of generator used for experiment}
\begin{center}
\begin{tabular}{|c|c|}
\hline
\textbf{Parameters} & \textbf{Values}\\
\hline
Line to Line Voltage & 400 V \\ \hline
Line to Neutral Voltage & 230 V \\ \hline
Line current & 2.6 A \\ \hline
Frequency & 50 Hz \\ \hline
Nominal Speed & 1500 rpm \\ \hline
Poles & 4 \\ \hline
Excitor current & 1.6 A \\ \hline
\end{tabular}
\label{tab1}
\end{center}
\end{table}
\section{Auto-Synchronization Design}
In this section, an auto-synchronization scheme has been proposed. The required apparatus are listed. Each step in the synchronization process is discussed.
\subsection{Apparatus and Equipment}
\begin{itemize}
\item Power meter module to read the required voltage and frequency of the Grid.
\item A synchronous generator with excitation system and prime mover.
\item Arduino UNO to give the control signals to excitation system of generator.
\item An 8 channel relay to amplify the signals by micro-controller.
\item An operating system to support the LabVEIW software.
\end{itemize}

\subsection{Frequency Synchronization}
Frequency control is the most important task regarding synchronous machine operation in real time. Using \eqref{eq}, one can get the value of the desired speed required by the synchronous generator to get synchronized with grid.
\begin{equation}
v=120f/P\label{eq}
\end{equation}
v is the speed of prime mover.\\
P is the number of poles of the synchronous generator.\\
f is the required frequency.\\

\paragraph{Speed Control Loop Design} 
The design requirement for frequency is about 50 Hz that corresponds to a speed of 1500 rpm with 4 poles. After obtaining the grid frequency and converting it into speed, the loop compares it with generator current speed. The loop keeps on monitoring the speed and adjusts the speed accordingly when required. 

\begin{figure}[]
\centerline{\includegraphics[width=8.6cm, height=3.6cm]{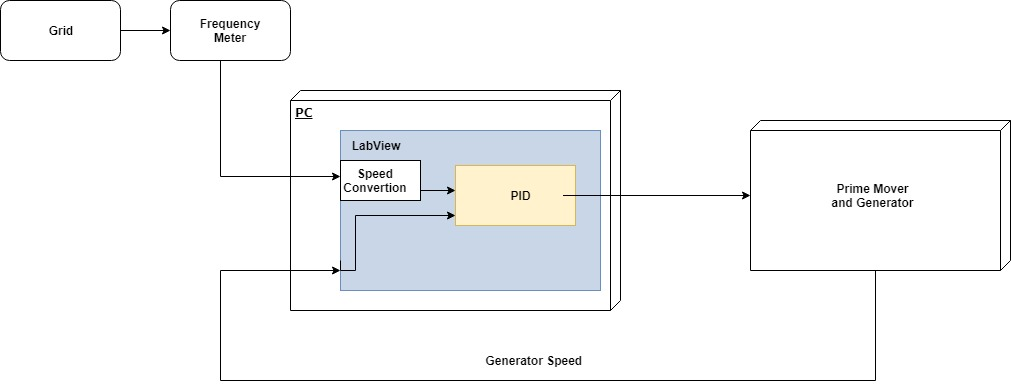}}
\caption{Block diagram for speed synchronization}
\label{fig2}
\end{figure}

\paragraph{Speed Control Loop Implementation} 
Speed Control is implemented using a proportional, integral and derivative controller (PID) controller in LabVIEW. Fig.~\ref{fig2} shows the block diagram representation of speed control. 

\begin{figure}[]
\centerline{\includegraphics[width=8.2cm, height=4cm]{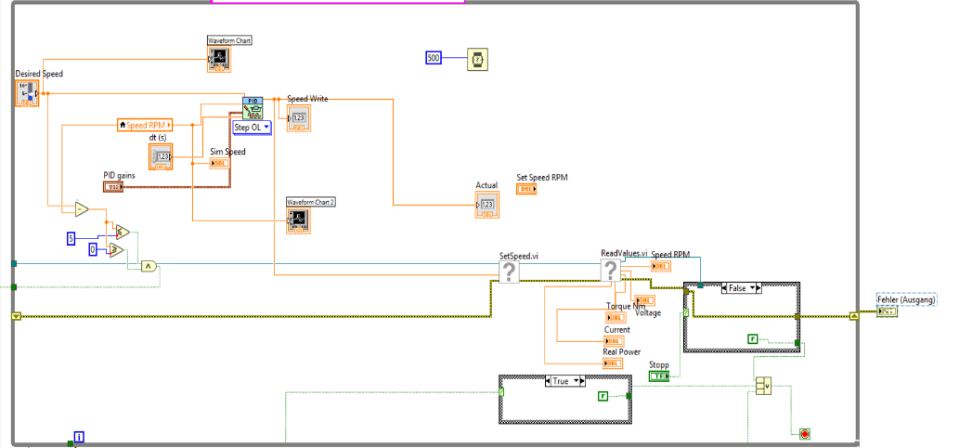}}
\caption{Block diagram for speed control in LabVIEW}
\label{fig3}
\end{figure}

\begin{figure}[]
\centerline{\includegraphics[width=0.5\textwidth, height=4.5cm]{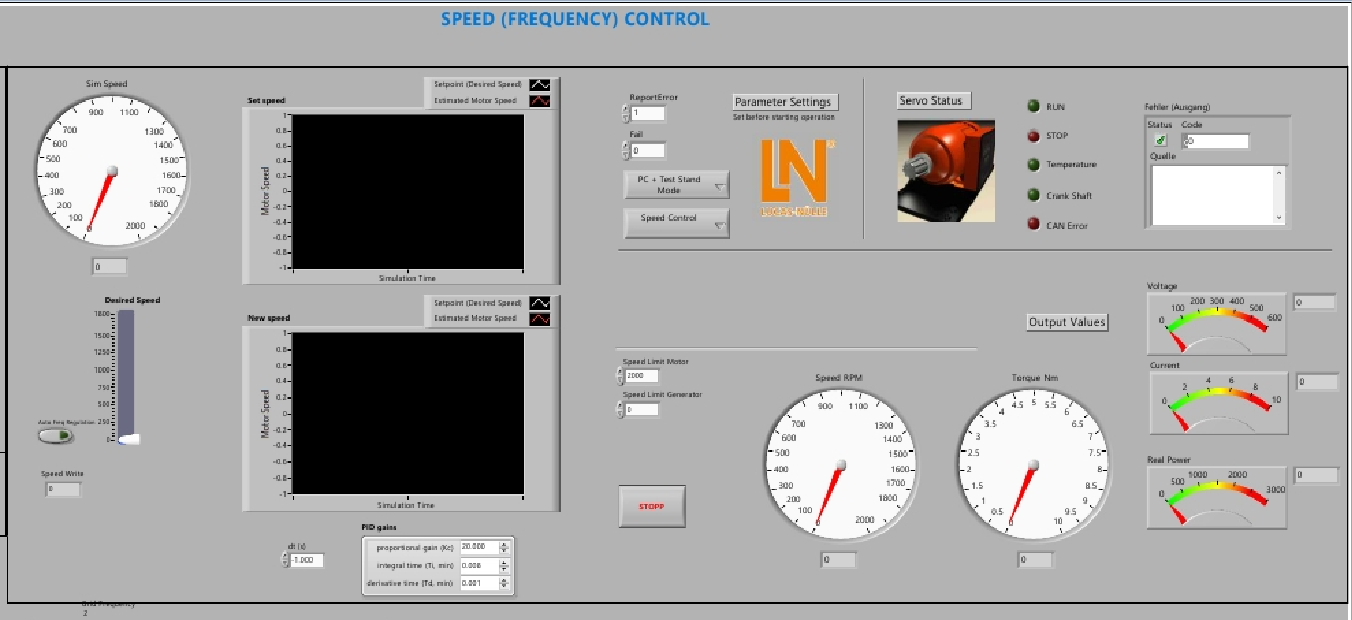}}
\caption{Front panel for speed control in LabVIEW}
\label{fig4}
\end{figure}

It is a control loop feedback system that constantly calculates the difference between a desired set speed and measured speed and applies a correction based on proportional, integral, and derivative terms to the generator on the basis of this difference. LabVIEW reads the values of voltages from grid and terminals of synchronous generator and uses PID controller to adjust the speed of synchronous generator. The response of PID is then fed to the prime mover, thus completing the feedback loop. Fig.~\ref{fig3} shows the block diagram and Fig.~\ref{fig4} shows the front panel window in LabVIEW of speed control loop.

\subsection{Voltage Synchronization}
\begin{figure}[]
\centerline{\includegraphics[width=0.3\textwidth]{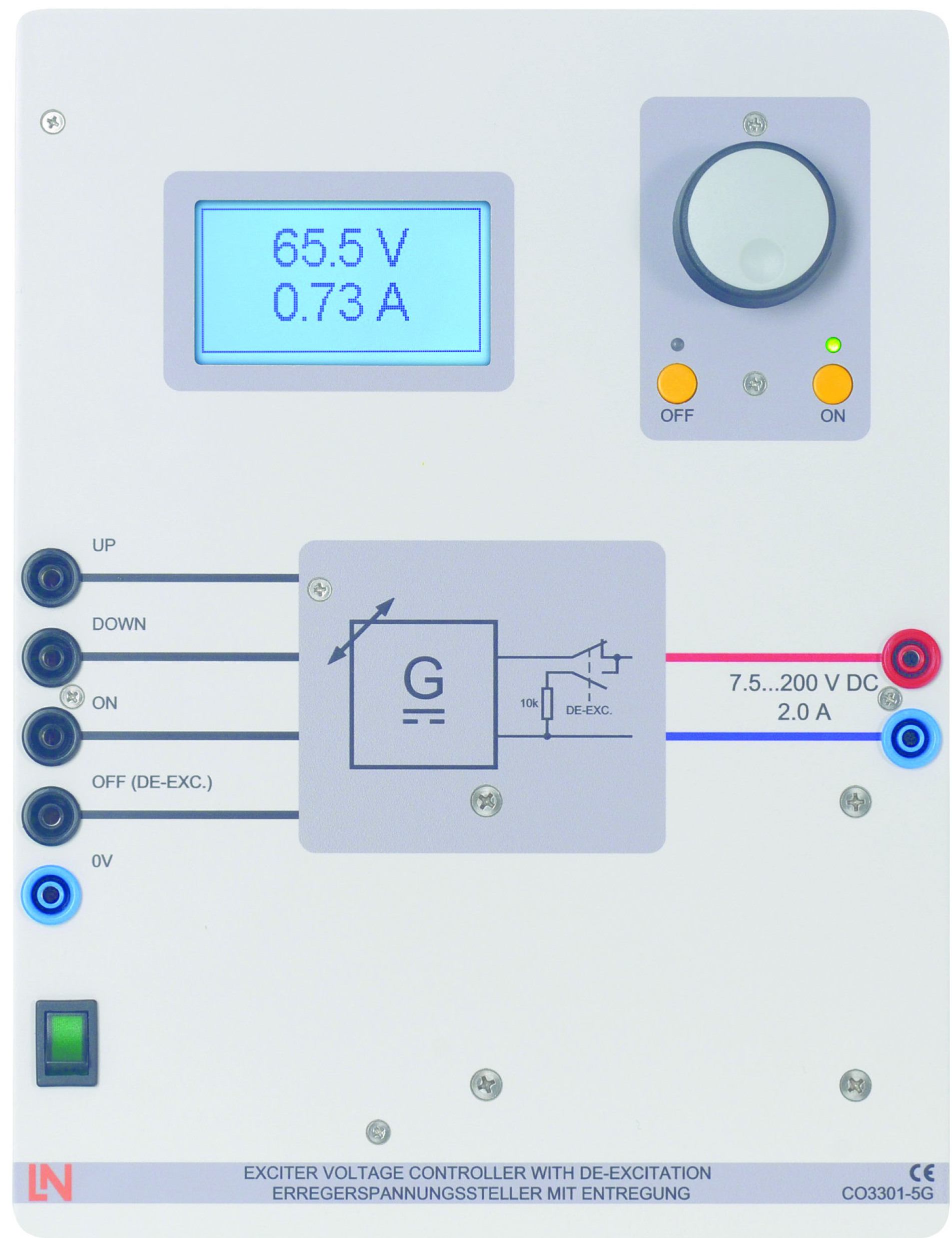}}
\caption{Excitation Module}
\label{exc}
\end{figure}

\paragraph{Voltage Control Loop Design} 
The output voltage of a synchronous generator can be regulated by controlling the excitation of the field voltage. The excitation voltage is provided by the Excitation Module Fig.~\ref{exc}. In the experimental setup of excitation system, there are 4 pins on the module. They are designed to switch ON the excitation module, to turn OFF the module, to increase the excitation voltage and decrease the excitation voltage.

\begin{figure}[]
\centerline{\includegraphics[width=8.59cm, height=3.69cm]{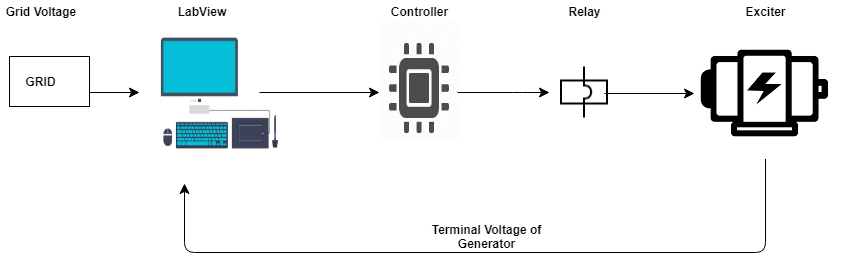}}
\caption{Voltage control loop block diagram}
\label{voltageLoop}
\end{figure}

To generate voltage control pulse signals Arduino UNO is used.
8-channel relay is used as an isolation relay, to provide a non-physical link between the microprocessor and the generator. It needs an input signal of 5V for its operation. The inputs from Arduino digital pins are connected to the switches of relay. To the output side, DC exciter system has been connected. For interfacing Arduino with LabVIEW we used LINX libary. Fig.~\ref{voltageLoop} shows the block diagram of voltage control loop.

\paragraph{Voltage Control Loop Implementation} 
Voltage Control is implemented by using a closed loop that obtains voltage of the grid and compare it with generator voltage. If the grid is at higher voltage, an UP pulse of 250 ms and 30 V magnitude is given to the exciter. Similarly, if the generator voltage exceeds the grid voltage, a DOWN pulse of 250 ms and 30 V magnitude is given to the exciter.

\begin{figure}[]
\centerline{\includegraphics[width=8cm, height=5.6cm]{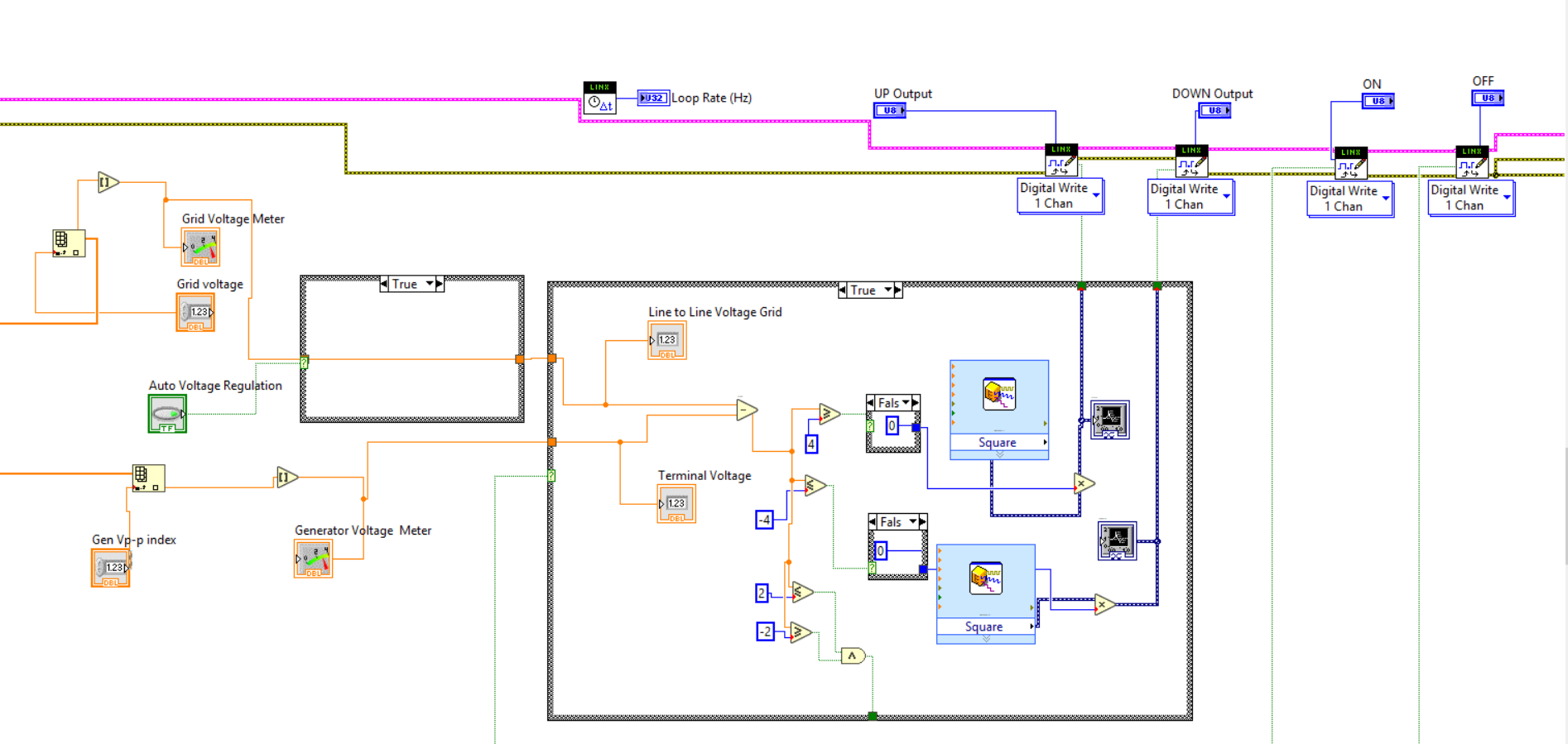}}
\caption{Block diagram for voltage control in LabVIEW}
\label{fig6}
\end{figure}

\begin{figure}[]
\centerline{\includegraphics[width=8cm, height=5.4cm]{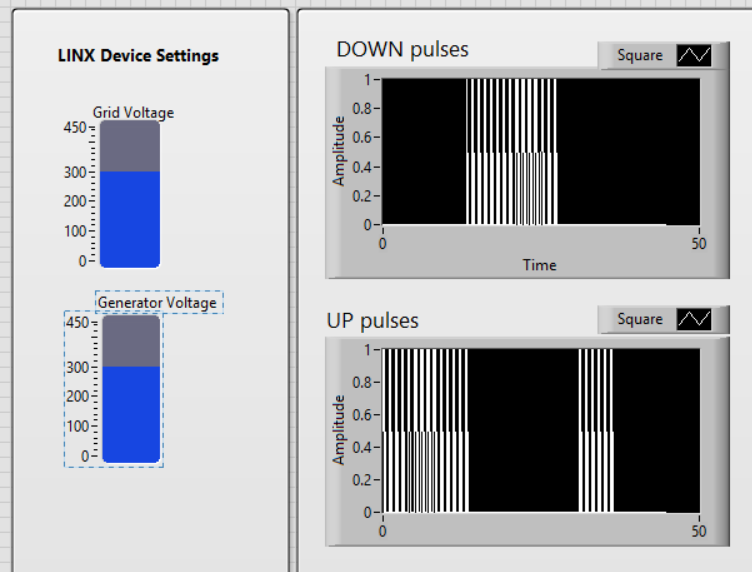}}
\caption{Front panel for voltage control in LabVIEW}
\label{fig7}
\end{figure}

On receiving the signals from LabVIEW, Arduino generates up and down digital signals at its digital pins according to the command. The 5V at the digital pins are converted to 30 V via the relay and then fed to the exciter system. The automatic voltage control loop keeps on monitoring the voltage and varies it accordingly, when required. Fig.~\ref{fig6} shows the block diagram and Fig.~\ref{fig7} shows the front panel window in LabVIEW of voltage control loop.

\subsection{Phase Synchronization}
\begin{figure}[]
\centerline{\includegraphics[width=8cm, height=5cm]{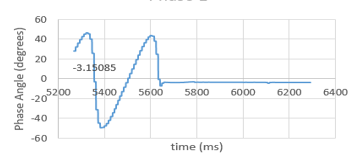}}
\caption{Phase difference curve with respect to time.}
\label{gra}
\end{figure}

The phase difference between the grid and the incoming generator needs to be ideally zero to remain in synchronism.
For phase synchronization, the switch gear needs to be closed at the time when the phase difference between the phase angle curve from the grid voltage and the phase angle curve from the upcoming generator voltage is touching the zero-reference line. Fig.~\ref{gra} shows phase difference curve that gradually attains value equal to zero.

The phase synchronization is attainable when the frequency of the connecting generator is matched with that of grid frequency. The phase angle will never get matched, if there is a difference in frequency.
So, the logic should check the time for which the phase difference remains zero.

\begin{figure}[]
\centerline{\includegraphics[width=0.5\textwidth]{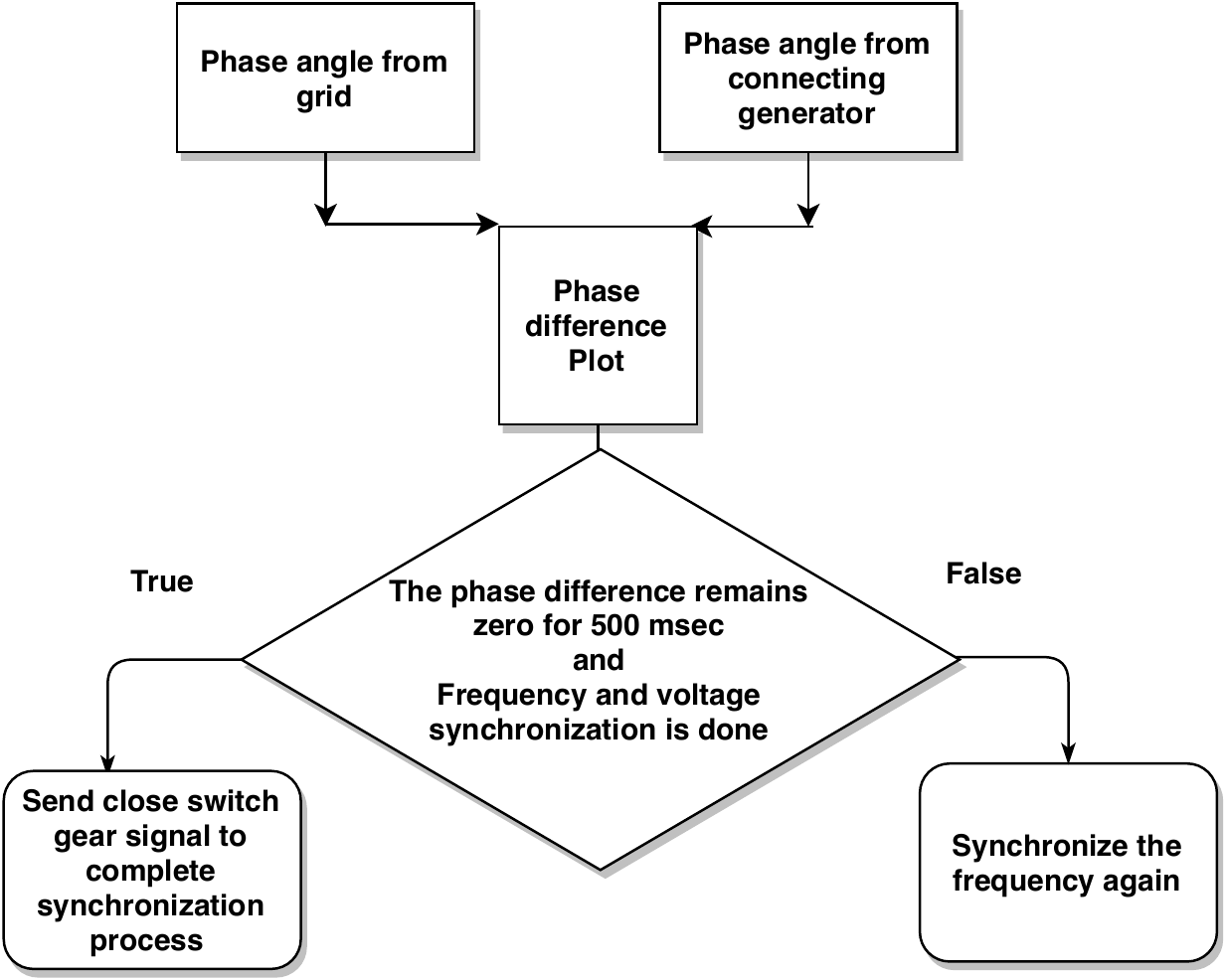}}
\caption{Phase synchronization block diagram.}
\label{fig8}
\end{figure}

When the voltage and frequency loops are implemented successfully, the system should check for the phase difference to be zero and then generates the signal to switch gear to close. Fig.~\ref{fig8} shows the block diagram of phase synchronization process.

\subsection{Flowchart of the proposed system}
\begin{figure}[]
\centerline{\includegraphics[width=0.5\textwidth]{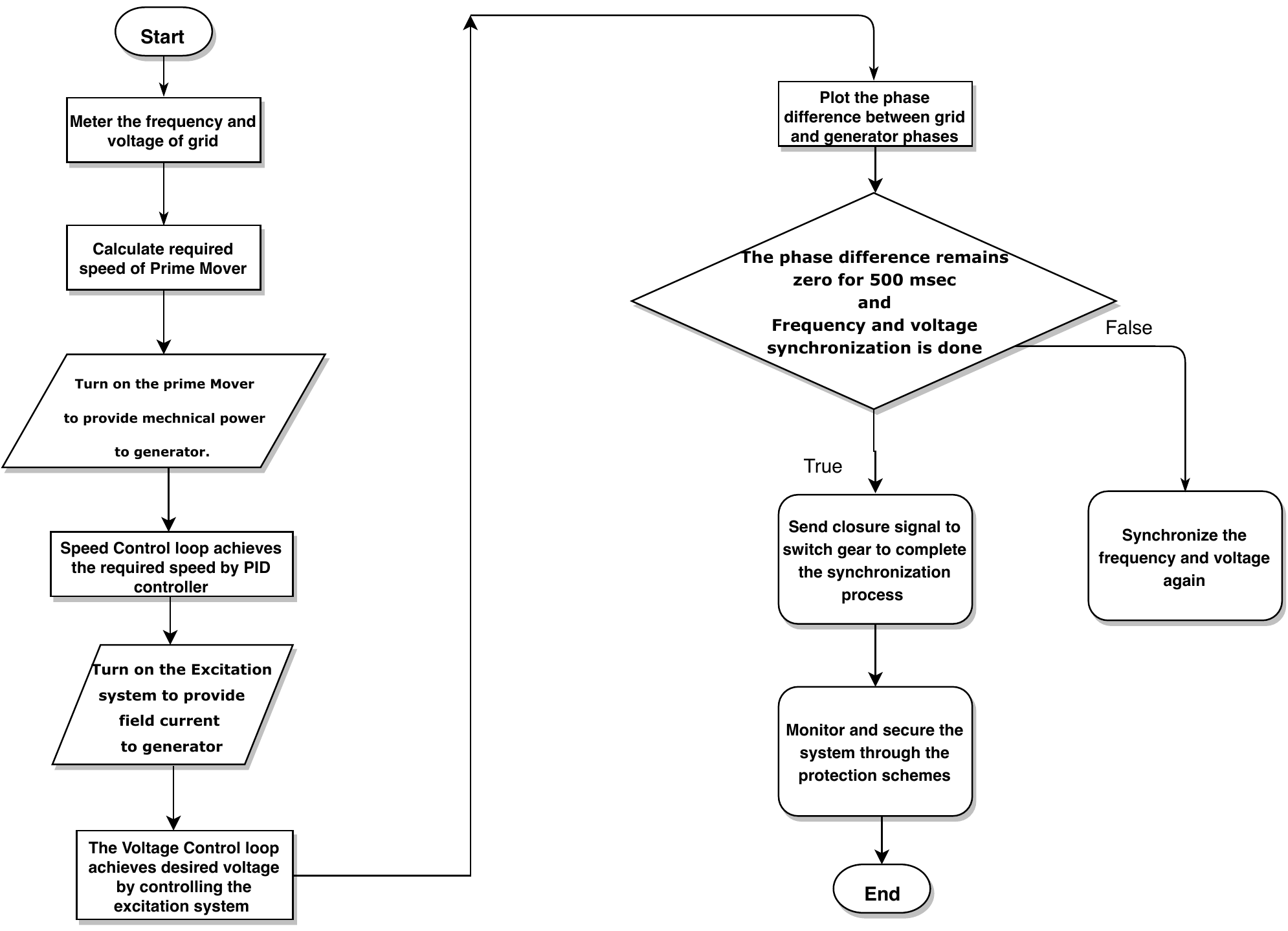}}
\caption{Flowchart of the  auto-synchronization process.}
\label{process}
\end{figure}

The proposed system monitors the grid parameters and adjust the speed, voltage and phase of the synchronous generator automatically. It also protects the generator against any faults. Fig.~\ref{process} shows the flowchart of the auto-synchronization process.

\section{Protection against damages and faults}
In order to avoid damage to the generator, it is ensured that the generator does not cross the limits of allowable range of frequency, current and voltage which can be dangerous  \cite{b12}. 
The alternator is taken out of service in case if control actions goes beyond the stability limits \cite{b13}.
Five protection schemes are designed to secure the generator.

\subsection{Overcurrent Protection Scheme} 
Overcurrent protection scheme interrupts the system in case a short circuit or a fault occurs. They can cause the current in the line to increase to a value much above the load current. 
When the current surpasses a predetermined value, the system is protected from the high current by the overcurrent relay \cite{b14}.

\subsection{Overvoltage Protection Scheme} 
Overvoltage protection scheme becomes compulsory for cases when there is a chance of the system shut down due to voltage exceeding to a hazardous value. So if the voltage level exceeds a preset level the overvoltage protection scheme trips the system and hence the system is saved from damage. 

\subsection{Undervoltage Protection Scheme} 
Undervoltage protection scheme operates in case when the line voltage becomes equal or less than a preset voltage.
The system is protected from under voltage which draws a large amount of current.

\subsection{Overfrequency Protection Scheme} 
Over frequency protection scheme uses a preset value of frequency and takes a threshold of a slightly higher value than 50Hz. If the system frequency gets higher than it, the circuit breakers trips. 

\subsection{Underfrequency Protection Scheme} 
\begin{figure}[]
\centerline{\includegraphics[width=7cm, height=9cm]{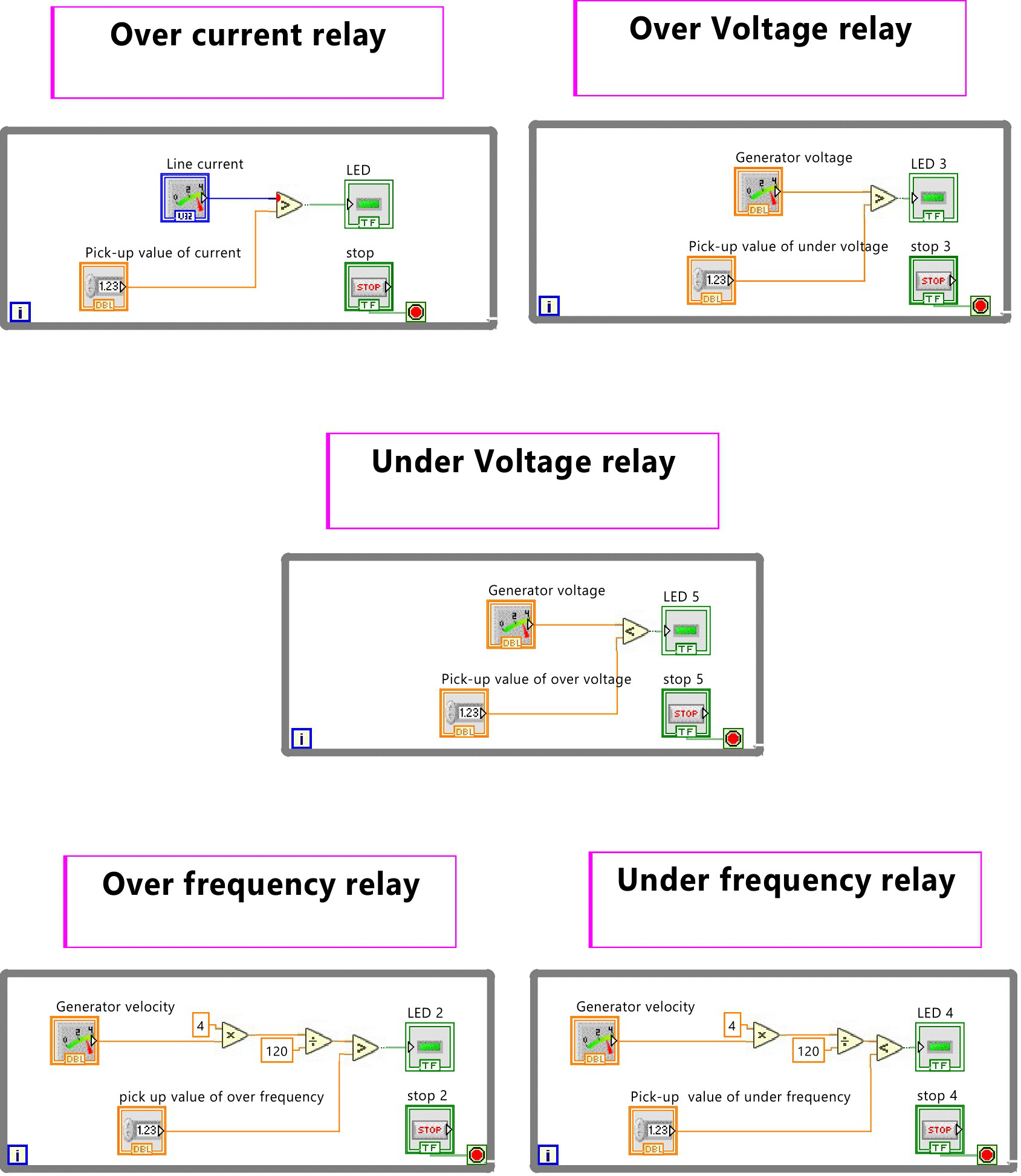}}
\caption{Protection scheme block diagram in LabVIEW.}
\label{block}
\end{figure}

\begin{figure}[]
\centerline{\includegraphics[width=7cm, height=8cm]{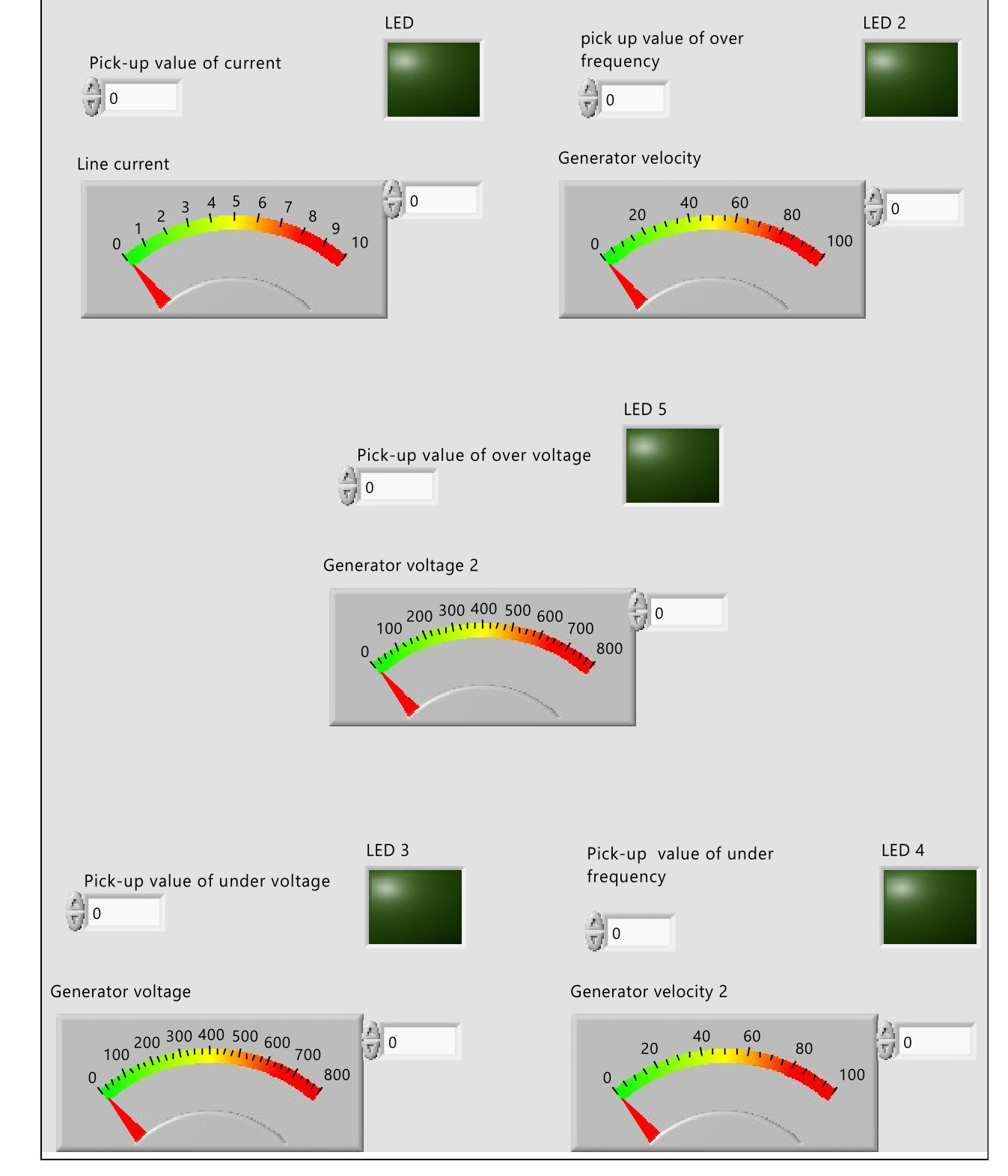}}
\caption{Protection scheme front panel in LabVIEW.}
\label{front}
\end{figure}

Under frequency protection scheme is important because even a slight decrease in frequency can have disastrous effects on the power system. So the system trips if frequency goes below 49.5Hz. \\
Fig.~\ref{block} shows the block diagram and Fig.~\ref{front} shows the front panel of protection scheme of generator in LabVIEW.

\section{Real Time Testing of Design}
All control loops were successfully designed and then merged into a single Virtual Instruments (VI) of LabVIEW.
All the connections were set up for the generator, the bus bars and the meters. 
Starting from the zero speed, the generator attained the synchronous speed of 1500 rpm corresponding to 50 Hz frequency and line to line voltage of 400 V at an excitation voltage of 45 V. Upon complete synchronization the system became ready to connect with grid. The simulation ran successfully.

\begin{figure}[]
\centerline{\includegraphics[width=7.8cm, height=5cm]{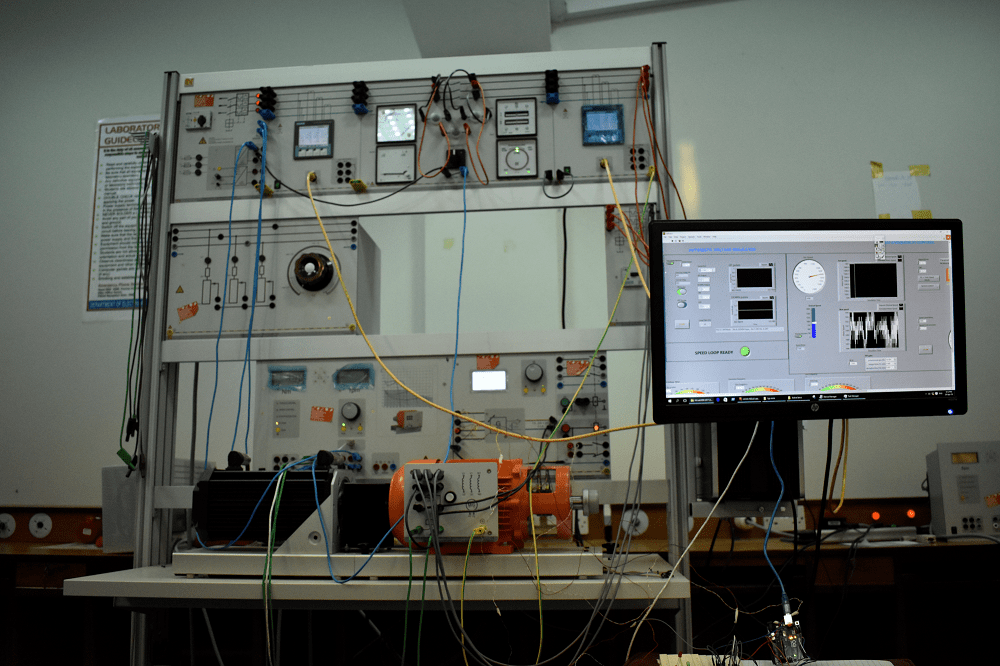}}
\caption{Experimental setup of working system.}
\label{setup}
\end{figure}

PHIL makes running and monitoring of the simulation in real time possible. The automatic system continuously keeps on monitoring the parameters and adjust them when required. The protection schemes keeps on monitoring and securing the system. 
 Fig.~\ref{setup} shows the experimental setup of the running system.

\section{Conclusions}
This paper has proposed an auto-synchronization relay which automates the process of synchronization of a Distributed Generator with the grid and makes the process easier, safer, cost-effective and reliable. In the software, custom auto-syncing modules have been developed successfully. A test bed system has been designed which is useful for the operation and control of a Distributed Generator in grid connected and isolated mode. The control loop for frequency synchronization of grid has been developed. The frequency of the connecting generator is adjusted by varying the speed of the prime mover, according to the frequency metered from the grid. Through the excitation control algorithm, an Automatic Voltage Regulation system has been developed to maintain desired voltage at generator's terminals. The safety and reliability of the whole scheme is ensured by damping any mechanical oscillations through power system stabilizer (PSS) and by providing the protection schemes. Protection schemes have been designed against over-current, under-voltage, over-voltage, over-frequency and under-frequency conditions the synchronous generator.

\end{document}